\documentclass{article}
\usepackage[utf8]{inputenc}
\usepackage{authblk}
\usepackage[dvips]{graphicx}
\graphicspath{{noiseimages/}}
\usepackage{xcolor}
\usepackage{float}
\usepackage[charter]{mathdesign}
\let\mathcal\undefined
\DeclareMathAlphabet{\mathcal}{OMS}{cmsy}{m}{n}
\usepackage[T1]{fontenc}
\usepackage{tikz}
\usepackage{empheq}

\usepackage{MnSymbol}
\usepackage{pgfplots}
\pgfplotsset{compat=1.7}
\usetikzlibrary{intersections, pgfplots.fillbetween}
\usetikzlibrary{decorations.pathmorphing}
\usetikzlibrary{arrows.meta}
\usepackage[mode=buildnew]{standalone}
\usepackage[toc,page]{appendix}
\usepackage{amsmath}
\usepackage{float}
\usepackage{comment}
\usepackage{bm}
\usepackage{a4wide}
\usepackage{bm}

\usepackage[english]{babel}
\usepackage{hyperref}
\usepackage{caption}
\AtBeginDocument{}
\definecolor{refcolor}{HTML}{CD2600}
\definecolor{tablecolor}{HTML}{373641} 
\definecolor{urlcolor}{HTML}{E12900} 
\hypersetup{pdfstartview=FitH,  linkcolor=refcolor,urlcolor=urlcolor, colorlinks=true,citecolor=refcolor}
\usepackage[page]{appendix}

\newcommand{\w}{\omega}

\author[1,2]{E. T. Akhmedov}
\author[1,2]{P. A. Anempodistov\footnote{\tt anempodistov.pa@phystech.edu}}
\author[1,2]{K. V. Bazarov\footnote{\tt bazarov.kv@phystech.edu}}
\affil[1]{Moscow Institute of Physics and Technology, Institutskii per. 9, 141700, Dolgoprudny, Russia}
\affil[2]{NRC ``Kurchatov Institute'', 123182, Moscow, Russia}
\title{\textcolor{black}{Nontrivial self-consistent backreaction of quantum fields in 2D dilaton gravity}}
\begin{document}

\numberwithin{equation}{section}

\maketitle

\begin{abstract}
    We consider $(1+1)$-dimensional dilatonic black hole with two horizons, canonical temperatures of which do not coincide. We show that the presence of quantum fields in such a background leads to a substantial backreaction on the metric: $(1+1)$-dimensional dilatonic analog of the semiclassical Einstein equations are solved self-consistently, and we demonstrate that taking into account of the backreaction leads to a geometry with two horizons with coinciding temperatures.
\end{abstract}

\newpage 
{
  \hypersetup{linkcolor=tablecolor}
}
\newpage\section{Introduction}



Spacetimes with multiple horizons are interesting backgrounds to consider from the perspective of consistency and/or stability when quantum fields are added. The notable example is the Schwarzschild-de Sitter spacetime. It is known that quantum fields in such a background in Euclidean signature inevitably lead to the conical singularity \cite{Kay:1988mu,Lin:1998pj,Teitelboim:2001skl,Shankaranarayanan:2003ya,Choudhury:2004ph}, whose presence poses a question of the consistency of quantum field theory in the backgrounds with multiple horizons. 

Here, however, we adopt a different logic. Namely, we assume that one can consider quantum fields in any state, which has a sensible stress energy tensor (e.g. conserved etc.) in backgrounds with Lorentzian signature. But if a quantum field is taken in a spacetime with a horizon e.g. in a thermal state which does not coincide with the canonical temperature (Unruh, Hawking or Gibbons-Hawking, correspondingly) then its regularized stress-energy tensor diverges at the horizon. We interpret such a situation as leading just to a strong backreaction on the geometry by the quantum fields (see e.g. \cite{Ho:2018jkm,Ho:2018fwq} and \cite{Anempodistov:2020oki,Bazarov:2021rrb,Akhmedov:2020ryq,Moretti:1997qn,PhysRevD.55.3552,PhysRevD.51.2770,Diatlyk:2020nxa,Diakonov:2023jdk}). 

Furthermore, if the geometry in question has multiple horizons, and the canonical temperatures of these horizons do not coincide, any thermal state of quantum fields leads to a divergence of the regularized stress-energy tensor on at least one of the horizons. Then in the situation with multiple horizons the regularized stress-energy tensor inevitably has a divergence, which causes strong backreaction on the background geometry. In this paper we consider a solvable example of 2D quantum field theory where such a conclusion can be checked. In this note we restrict our attention to the case when quantum fields are considered on a classical gravitational background.

In \cite{Markovic:1991ua} the authors have considered the two-dimensional analogue of Schwarzschild-de Sitter spacetime, and argued that for the thermal states the stress-energy tensor diverges at least on one of the horizons. They also conclude that the divergence of stress-energy tensor signals that the state of quantum fields is unphysical. Then, they attempt to construct a state which is regular on both horizons. We, however, adopt a different logic, that we have mentioned above and which is in accordance with \cite{Akhmedov:2022qpu}. The point is that thermal states in stationary backgrounds have a notable advantage in comparison with other states: according to the fluctuation-dissipation theorem they are stable under quantum corrections for self-interacting fields\footnote{In \cite{Markovic:1991ua} they consider a gaussian theory in the background in question. In this note we also restrict our attention to the gaussian theory, but keep in mind effects of self-interactions. Also in two dimensions the situation with the fluctuation-dissipation theorem is different from higher dimensions due to the kinematic reasons, but we consider 2D case as the model example.}.

The problem of backreaction of matter fields on background geometries has been addressed by many authors (see e.g. \cite{Ho:2018jkm, Ho:2018fwq, Baake:2023gxx} for the related works). In those papers the Einstein equations
\begin{align}
    G^{\,\,\,\mu}_{\nu}+\Lambda \delta^{\,\,\,\mu}_{\nu}=8\pi G  \langle :\hat{T}^{\,\,\,\mu}_{\nu}:\rangle,\label{ein}
\end{align}
are solved perturbatively:
\begin{itemize}
    \item First, solution $g^{(0)}$ of the Einstein equations in the vacuum (i.e. with the zero on the r.h.s.) is considered (usually a black hole with $\Lambda = 0$);
    \item Second, the stress-energy tensor for quantum fields in the background $g^{(0)}$ is calculated;
    \item Third, the Einstein equations are solved perturbatively for a new metric $g$ on the l.h.s., while on the r.h.s. one plugs the stress-energy tensor calculated with the metric $g^{(0)}$. 
\end{itemize}
In this approach, the new metric $g$ is usually divergent at the position of the horizon. 

In our paper we take a different approach. Namely, we solve the 2D analog of Einstein equations (in dilatonic gravity theory) self-consistently, i.e. the stress-energy tensor is calculated for the same metric which is plugged on the l.h.s. of these equations. Of course, for this method to work, one has to know the form of the expectation value of the stress-energy tensor in a generic background. That is the reason why at this stage we have to confine our analysis to the 2D situation, where the exact form of the regularized stress-energy tensor is known at any point in spacetime for arbitrary metric. 

Note that there is another method to tackle the backreaction problem using the effective action approach (see, e.g., \cite{Giddings:1992fp, Callan:1992rs, deAlwis:1992hv, Russo:1992ax, Russo:1992yh, Fitkevich:2020okl, Solodukhin:1995te, Potaux:2021yan, Sadekov:2021rpy, Svesko:2022txo}). The advantage of our method is that we can explicitly calculate the quantum average of stress-tensor for any level population (see \eqref{T00n}).

Usually to make the gravity in 2D dynamical one has to add the dilaton field:
\begin{align}
    S^{\text{grav}}= \frac{1}{16\pi G}\int d^2x \sqrt{-g} e^{-2\phi}\Big[R-4\omega (\partial_\mu \phi )^2+4\lambda^2\Big],
\end{align}
where $\w$ is an arbitrary constant. For particular values of this $\w$ parameter this action reduces to the known theories:
\begin{itemize}
    \item $\omega=0$ case is the Jackiw-Teitelboim theory;
    \item if $\omega=-\frac{1}{2}$ one obtains planar general relativity;
    \item $\omega=-1$ one has the first-order string theory.
\end{itemize}
To make the kinetic term canonical one can always redefine the dilaton field, but then the dilaton coupling is modified.

The equations of motion following from the variation of this action with respect to the metric and dilaton field, correspondingly, are as follows:
\begin{align} \label{set_grav}
    e^{2\phi}T_{\mu\nu}^{\text{grav}} \equiv -2(\omega+1)D_\mu \phi D_\nu \phi+D_\mu D_\nu \phi-g_{\mu\nu} D_\mu D^\nu \phi+(\omega+2)g_{\mu\nu}D_\mu \phi D^{\mu}\phi-g_{\mu\nu}\lambda^2=0,
\end{align}
\begin{align}
\label{EOMphi}
    R-4\omega D_\mu D^\mu \phi+4\omega D_\mu \phi D^{\mu}\phi+4\lambda^2=0.
\end{align}
The comprehensive analysis of solutions of these equations of motion for arbitrary $\w$ can be found in \cite{Lemos:1993py} (see \cite{Nojiri:1998yg, Nojiri:1999br, Nojiri:2000ja, vanNieuwenhuizen:1999nu, Cadoni:2023tse} for a review of this model as well as discussion of various quantum effects). For a particular choice of the integration constant for the dilaton, the solution of these equations of motion can be represented in the following form:
\begin{align} \label{sol}
    &\phi = - \log(ar)^{\frac{1}{2(\w+1)}}, \nonumber \\
    &ds^2 = -[a^2r^2 - (ar)^{\frac{\w}{\w+1}}] dt^2 + \frac{dr^2}{a^2r^2 - (ar)^{\frac{\w}{\w+1}}},
\end{align}
for some $a$ related to $\lambda$ and to be defined below.

We begin in Sec. \ref{sec2} by discussing the properties of this solution for the case when $\w = - \frac{4}{3}$. The most important property of such a solution is the presence of two horizons with different temperatures. Then, in Sec. \ref{sec3} we add gaussian quantum fields that live in the background of this geometry, and demonstrate that the regularized stress-energy tensor diverges at least on one of the horizons. In Sec. \ref{sec4} we solve semiclassical gravitational equations and show that the resulting backreacted geometry has two horizons with coinciding temperatures. 

\section{Starting geometry} \label{sec2}

\begin{figure}[H]
\centering
\includegraphics[width=0.7\textwidth]{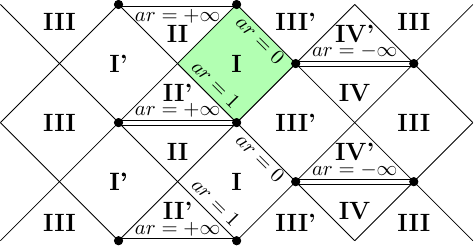}
\caption{Maximal analytical extension of the metric \eqref{metric}. \textcolor{black}{There are infinite copies of this diagram extending both in vertical and horizontal direction. Coordinates \eqref{metric} cover only chart \textbf{I} (highlighted by green on the diagram). The boundary of the chart are lightlike surfaces corresponding to $ar=0$ and $ar=1$}.}
\label{penrose}
\end{figure}

Solutions (\ref{sol}) with different $\omega$ correspond to different physical situations, and we want to consider the most similar one to the black hole in de Sitter space-time. We are looking for multiple horizon scenario, where horizons have different temperatures (surface gravity). To model this situation, we propose to consider the case $\w=-\frac{4}{3}$. Then the metric takes the form:
\begin{equation} \label{metric}
    ds^2 = -[a^2r^2 - a^4r^4] dt^2 + \frac{dr^2}{a^2r^2 - a^4r^4},
\end{equation}
while the dilaton field acquires the form:
\begin{align}
    \phi=\frac{3}{2}\log(ar),
\end{align}
where $a=-\sqrt{\frac{2\lambda^2}{3}}$. The Ricci scalar for the metric in question is $R = -\frac{4}{3} \lambda^2 \left[1-6(ar)^2\right]$. This metric is defined for $r>0$ and possesses two horizons: at $r=0$ and at $r=1/a$. The canonical temperatures at the horizons are:
\begin{equation}
    T = \begin{cases}
        \frac{a}{2\pi}, \qquad r \to 1/a,\\
        0, \,\,\,\, \qquad r \to 0.
    \end{cases}
\end{equation}
Note that the zero of the function $[a^2r^2 - a^4r^4]$ at $r=0$ is degenerate. As a result, the corresponding temperature is vanishing. The Penrose-Carter diagram for this spacetime is depicted on the Fig.\ref{penrose}.
For such space-times as we consider it is useful to use other (tortoise) coordinates: 
\begin{equation} 
    ds^2 = -\left[a^2r^2 - a^4r^4\right] \, \left( dt^2 - dr_*^2\right),
\end{equation}
where 
\begin{equation}
    r_* = -\frac{1}{a^2 r} + \frac{1}{a} \text{arctanh}(ar).
\end{equation}
Or using Eddington-Finkelstein coordinates
\begin{equation}
    u=t-r_*, \qquad v=t+r_*,
\end{equation}
we can rewrite the metric in the form:
\begin{equation} \label{metric_conf}
    ds^2 = C(u,v) du dv, \qquad C(u,v) = -a^2r^2(u,v) + a^4r^4(u,v),
\end{equation}
where $r(u,v)$ is determined from
\begin{equation}
    \frac{v-u}{2} =  -\frac{1}{a^2 r} + \frac{1}{a} \text{arctanh}(ar).
\end{equation}
Using substitution
\begin{align} \label{r_to_x}
    r=\frac{1}{a \cosh(a X)},
\end{align}
we can write the metric in the form
\begin{align}
\label{nucoordinates}
    ds^2=-e^{2\nu(X)}dt^2+dX^2, 
\end{align}
where
\begin{equation} \label{nu0}
    e^{2\nu(X)} =\frac{\sinh^2(aX)}{\cosh^4(aX)}.
\end{equation}
In these coordinates, the dilaton field is equal to
\begin{equation} \label{phi0}
    \phi(X) = -\frac{3}{2} \log \cosh(aX).
\end{equation}
In this note we use all these coordinate systems in different situations, depending on where it is convenient to use one of them.

\section{Adding quantum fields} \label{sec3}

The next step in our reasoning is to add quantum fields. We consider real conformal field $\varphi$ with the action:
\begin{align}
    S=S^{\text{grav}}+S^{\text{matter}}=S^{\text{grav}}-\frac{1}{2}\int d^2x\partial_\mu \varphi \partial^\mu \varphi,
\end{align}
just as in e.g. \cite{Giddings:1992fp, Callan:1992rs, deAlwis:1992hv, Russo:1992ax, Russo:1992yh}, but rather than integrating the field $\varphi$ out we explicitly quantize it and calculate the regularized stress-energy tensor. We will adopt the canonical quantization.

It is useful to use the Eddington-Finkelstein coordinates \eqref{metric_conf}, in which the equations of motion for the scalar field take the form:
\begin{align}
    \partial_U \partial_V \varphi(U,V)=0.
\end{align}
Thus, we can write down the mode decomposition of the field operator as:
\begin{align}
    \hat{\varphi}(U,V)=\int_0^{\infty}\frac{d\omega}{\sqrt{4\pi\omega}}\Big[\hat{a}^\dagger_\omega e^{i\omega U}+\hat{b}^\dagger_\omega e^{i\omega V}+\text{h.c.}\Big].
\end{align}
Commutation relations for the creation and annihilation operators for the left and right moving modes are:
\begin{align}
[\hat{a}^\dagger_\omega,\hat{a}^{}_{\omega'}]=\delta(\omega-\omega'), \quad [\hat{b}^\dagger_\omega,\hat{b}^{}_{\omega'}]=\delta(\omega-\omega').
\end{align}
We take the quantum field in the thermal state, for which:
\begin{align} \label{state}
     \langle \hat{a}_\w^{\dagger} \hat{a}^{}_{\w'}\rangle=\langle \hat{b}_\w^{\dagger} \hat{b}^{}_{\w'}\rangle=\frac{1}{e^{\beta \w}-1}\delta(\w-\w'),
\end{align}
where $\beta = 1/T$ is the inverse temperature.

Then, the regularized stress-energy tensor is given by \cite{Birrell:1982ix}:
\begin{equation} \label{set_m}
    - \langle : T_{\mu \nu}^{\text{matter}} : \rangle = \Theta_{\mu \nu} + \frac{R}{48\pi} g_{\mu \nu},
\end{equation}
where
\begin{align}
    \Theta_{uu} &= \frac{1}{48\pi} \bigg( \frac{2\pi}{\beta}  \bigg)^2 -\frac{1}{12 \pi} C^{1/2} \partial^2_u C^{-1/2}, \nonumber \\
    \Theta_{vv} &= \frac{1}{48\pi} \bigg( \frac{2\pi}{\beta}  \bigg)^2 -\frac{1}{12 \pi} C^{1/2} \partial^2_v C^{-1/2} , \\
    \Theta_{uv} &= \Theta_{vu} =0,\nonumber
\end{align}
and $C(u,v)$ here is equal to \eqref{metric_conf}.
Then, plugging the expression for the conformal factor, $C(u,v)$, we find that
\begin{equation}
    \Theta_{uu} = \begin{cases}
    \frac{1}{48\pi} \bigg[ \big( \frac{2\pi}{\beta}  \big)^2 - a^2   \bigg], \qquad \text{as} \quad r \to \frac{1}{a} ,  \\
    \frac{1}{48\pi}  \big( \frac{2\pi}{\beta}  \big)^2, \qquad \qquad \quad \text{as} \quad r \to 0,
    \end{cases} \label{T_two_hor}
\end{equation}
and the same expression for $\Theta_{vv}$. We see that one can choose the inverse temperature $\beta$ such that $\Theta_{\mu \nu}$ is vanishing only on one of the horizons. If $\Theta_{\mu \nu}$ has a non-zero value on a horizon, it implies that $T^{\mu}_{\nu}$ and $T^{\mu \nu}$ blow up on this horizon, which leads to the substantial backreaction on the geometry.

In the coordinates \eqref{nucoordinates} the stress energy tensor has the following form\footnote{Note, that the stress energy tensor is conserved $D_\mu \langle : T^{\mu\nu}_{\text{matter}}:\rangle =0$.}:
\begin{align}
\label{Ttt}
    -\langle : &T_{tt}^{\text{matter}}:\rangle =\frac{\pi}{6\beta^2}+\frac{e^{2\nu(X)}}{24\pi} \Big[\nu'(X)^2+2\nu''(X)\Big], \nonumber \\
    -\langle : &T_{XX}^{\text{matter}}:\rangle =\frac{e^{-2\nu(X)}\pi}{6\beta^2}-\frac{\nu'(X)^2}{24\pi}.
\end{align}
Below we will use these expressions to find a self-consistent solution of the modified equations of motion.

We consider only thermal states in our calculations. Such a state has level population given by the planckian distribution. However, our method allows one to extend the calculation for any level population. For example, for generic level population $n(\omega)$ energy density is given by the following expression:
\begin{align}
\label{T00n}
    -\langle : &T_{tt}^{\text{matter}}:\rangle =\int_{-\infty}^{\infty} \frac{\omega d\omega}{2\pi} \Big[n(\omega)+\Theta(-\omega)\Big]+\frac{e^{2\nu(X)}}{24\pi} \Big[\nu'(X)^2+2\nu''(X)\Big],
\end{align}
where 
\begin{align} 
     \langle \hat{a}_\w^{\dagger} \hat{a}^{}_{\w'}\rangle=\langle \hat{b}_\w^{\dagger} \hat{b}^{}_{\w'}\rangle=n(\omega)\delta(\w-\w'),
\end{align}
and $\Theta(-\omega)$ is the Heaviside step function. If $n(\omega)=[e^{\beta\omega}-1]^{-1}$, then:
\begin{align}
    \int_{-\infty}^{\infty} \frac{\omega d\omega}{2\pi} \Big[\frac{1}{e^{\beta\omega}-1}+\Theta(-\omega)\Big]=\frac{\pi}{6\beta^2}
\end{align}
and we obtain \eqref{Ttt}.

\section{Backreaction of quantum fields on the background geometry} \label{sec4}

Now, let us continue with the solution of the 2D dilatonic analog of the semiclassical Einstein equations (\ref{ein})
self-consistently, i.e. by plugging in it an arbitrary metric and regularized stress-energy tensor of the matter fields calculated in the background of the same metric. In the 2D dilaton gravity that we consider here these equations of course look differently from (\ref{ein}).

Let us start with writing down non-zero components of the gravitational stress energy tensor in the unitary gauge \eqref{nucoordinates}\footnote{Note, that the third equation $\frac{\delta S^{\text{grav}}}{\delta \phi(X)}=0$ is satisfied if $T^{\text{grav}}_{\mu\nu}=8\pi G \langle :T^{\text{matter}}_{\mu\nu}:\rangle$.}:
\begin{align}
&e^{2\phi(X)}T^{\text{grav}}_{tt}=e^{2\nu(X)}\big[\lambda^2-\frac{2}{3}\phi'(X)^2+\phi''(X)\big], \nonumber \\   &e^{2\phi(X)}T^{\text{grav}}_{XX}=-\lambda^2-\phi'(X)\nu'(X)+\frac{4}{3} \phi'(X)^2.
\end{align}
Then, the 2D dilatonic analog of Einstein equations \eqref{ein} reduces to:
\begin{align}
    &\lambda^2-\frac{2}{3}\phi'(X)^2+\phi''(X) =8\pi G e^{2\phi(X)}\bigg(\frac{e^{-2\nu(X)}\pi}{6\beta^2}+\frac{1}{24\pi} \Big[\nu'(X)^2+2\nu''(X)\Big]\bigg), \nonumber \\
    &-\lambda^2-\phi'(X)\nu'(X)+\frac{4}{3} \phi'(X)^2=8\pi G e^{2\phi(X)}\bigg(\frac{e^{-2\nu(X)}\pi}{6\beta^2}-\frac{\nu'(X)^2}{24\pi}\bigg), \label{back_eq}
\end{align}
where now $\phi(X)$ and $\nu(X)$ are unknown functions to be determined from these equations. 

When $\beta = 0$ (the case of interest for us, as is explained below), these equations have an obvious solution with $\nu(X)$ being a linear function of $X$ and $\phi$ being a constant. The constants in the solution should be suitably adjusted according to the values of $\lambda$ and $G$. 
However, that is not a solution that we are looking for. Because our solution should have an appropriate asymptotic behavior at one of the horizons: we want to find what happens under backreaction (due to the quantum fields) with the solution with two horizons. 

\begin{figure}[H]
\centering
\includegraphics[width=1.0\textwidth]{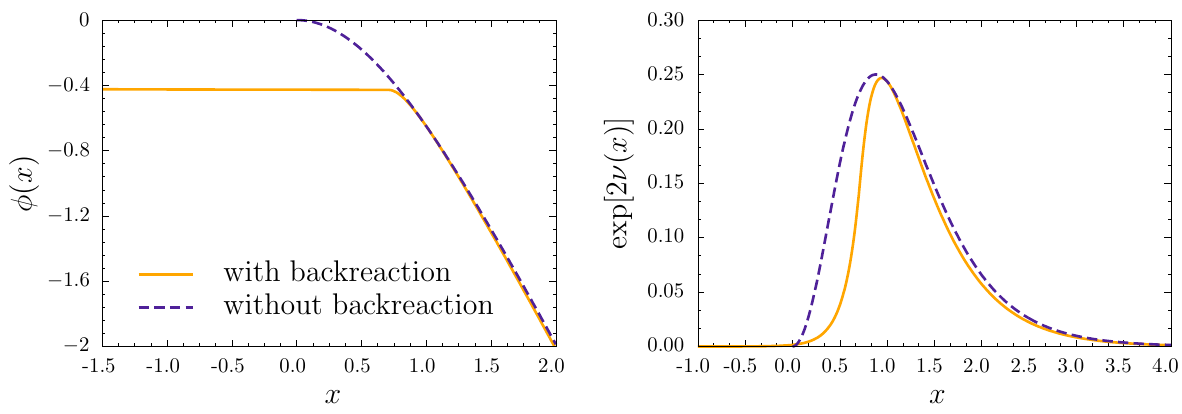}
\caption{Numerical plot of the dilaton field $\phi(x)$ and for the $e^{2\nu(x)}$ with and without account of backreaction. Note that the solution without backreaction is defined only for $X>0$, while solution with the backreaction is defined on the whole $X$ axis.}
\label{fig}
\end{figure}

To find the solution of \eqref{back_eq} with appropriate asymptotic behavior we have to apply numerical methods. To do that, we need to specify boundary conditions. We will assume, that the horizon at $r=0$ ($X=\infty$) remains unchanged. Comparing with \eqref{T_two_hor}, one can see that this conditions implies that $1/\beta = 0$ (in other words, the state of the field is the Fock space ground state). Hence, Eq. \eqref{back_eq} for the dimensionless variable $x=|a|X$ (with $a = -\sqrt{\frac{2\lambda^2}{3}}$) reads\footnote{Here we have set $G=1$.}:
\begin{align}
    &\frac{3}{2}-\frac{2}{3}\phi'(x)^2+\phi''(x) =\frac{e^{2\phi(x)}}{3}  \Big[\nu'(x)^2+2\nu''(x)\Big], \nonumber \\
    &-\frac{3}{2}-\phi'(x)\nu'(x)+\frac{4}{3} \phi'(x)^2=-\frac{e^{2\phi(x)}}{3} \nu'(x)^2. \label{back_eq_dimensionless}
\end{align}
The solution of these equations for the dilaton field and for the $e^{2\nu(x)}$ have the form shown on the Fig. \ref{fig}.
These forms of the solutions could have been guessed from the following perspective: the solutions of \eqref{back_eq_dimensionless} should interpolate from the form of \eqref{phi0} and \eqref{nu0} at $x \gg x_0$ (where $x_0$ is a reference gluing point) to the solution
\begin{equation} \label{linear_sol}
    \phi(x) \approx \phi_0 = \text{const}, \qquad \nu(x) \approx \frac{3}{\sqrt{2}} e^{-\phi_0} x +\text{const},
\end{equation}
at $x \ll x_0$, as follows from \eqref{back_eq_dimensionless}. As can be seen from the Fig. \ref{fig} these are precisely solutions obeying such boundary conditions.

There is one important point to note here. The solutions \eqref{phi0} and \eqref{nu0} are defined in the region $X>0$ (from single-valuedness of \eqref{r_to_x}), but the backreacted geometry of the Fig. \ref{fig} is defined on the entire axis $X \in (-\infty;+\infty)$. The new geometry has two horizons at $X=\pm \infty$, and canonical temperatures at these two horizons are vanishing\footnote{This can be seen from the examination of the contribution $C^{1/2} \partial^2_u C^{-1/2}$ to the stress energy tensor at $X=\pm \infty$.}. Thus, the backreacted geometry with the quantum state under consideration is in fact stable. 

\section{Conclusions}

We have started with the geometry that possessed two horizons with different canonical temperatures. The fact that the two horizons have different temperatures leads to substantial backreaction of quantum fields on the background metric -- the divergence of the stress-energy tensor on one of the horizons does not allow us to neglect the influence of quantum fields on the metric of the spacetime at this horizon. This happens if one considers fields to be in a thermal state, and one could, in principle, define a very specific state for which the regularized stress-energy tensor is regular on both horizons (see e.g. \cite{Markovic:1991ua}). But unlike the thermal one any another state will not necessarily survive quantum corrections in self-interacting theory (at least in dimensions higher than two).

Then, we have written down 2D dilatonic analog of the semiclassical Einstein equations in a self-consistent manner, i.e. the stress-energy tensor on the r.h.s. of these equations was calculated for arbitrary metric, which is to be determined from the resulting equations. 
Then, we have solved these equations numerically with the boundary condition that one of the horizon is unchanged.

We have found quite a remarkable result that the account of the backreaction changes the geometry in a way that the two horizons have the same (vanishing) temperature. Hence, after the introduction of quantum fields and accounting of the backreaction we obtain a geometry that is stable for the vacuum state (the thermal state with vanishing temperature) of the quantum field. 

\section{Acknowledgements}

We would like to thank Artem Alexandrov for valuable discussions regarding numerical solution of equations.
This work
was supported by the Foundation for the Advancement
of Theoretical Physics and Mathematics “BASIS” and by Russian Ministry of education and science.




\bibliographystyle{unsrturl}
\bibliography{bibliography.bib}
\end{document}